\documentclass[sigconf]{acmart}

\usepackage{booktabs}
\usepackage{xcolor}
\usepackage{graphicx}
\usepackage{multirow}
\usepackage{subcaption}
\usepackage[ruled,vlined]{algorithm2e}
\graphicspath{{figures/}}
\usepackage{amsmath}
\usepackage{verbatim}
\usepackage{threeparttable}
\usepackage{bbm}

\newcommand{\system}{AutoFD\xspace}

\def\BibTeX{{\rm B\kern-.05em{\sc i\kern-.025em b}\kern-.08emT\kern-.1667em\lower.7ex\hbox{E}\kern-.125emX}}
    
\copyrightyear{2018}
\acmYear{2018}

\newcommand\ci{\perp\!\!\!\perp}
\renewcommand\footnotetextcopyrightpermission[1]{} 
\begin{document}

\begin{abstract}
We study the problem of discovering functional dependencies (FD) from a noisy dataset. We focus on FDs that correspond to statistical dependencies in a dataset and draw connections between FD discovery and structure learning in probabilistic graphical models. We show that discovering FDs from a noisy dataset is equivalent to learning the structure of a graphical model over binary random variables, where each random variable corresponds to a functional of the dataset attributes. We build upon this observation to introduce \system a conceptually simple framework in which learning functional dependencies corresponds to solving a sparse regression problem. We show that our methods can recover true functional dependencies across a diverse array of real-world and synthetic datasets, even in the presence of noisy or missing data. We find that \system scales to large data instances with millions of tuples and hundreds of attributes while it yields an average $F_1$ improvement of $2\times$ against state-of-the-art FD discovery methods.
\end{abstract}

\title{Learning Functional Dependencies with Sparse Regression}

\author{Zhihan Guo, Theodoros Rekatsinas}

\maketitle

\section{Introduction}\label{sec:intro}
Functional dependencies (FDs) are an integral part of data management systems. They are used in database normalization to reduce data redundancy and improve data integrity~\cite{Garcia-Molina:1999:DSI:553977}. FDs are also critical in data preparation tasks, such as data profiling and data cleaning. For instance, FDs can help guide feature engineering in machine learning pipelines~\cite{PhysRevLett.114.105503} or can serve as a means to identify and repair erroneous values in the given dataset~\cite{hc, holistic}. Unfortunately, FDs are typically unknown and significant effort and domain expertise are required to identify them.

Various works have focused on automating FD discovery, both in the database~\cite{kruse2018efficient, tane, Papenbrock:2015:FDD:2794367.2794377} and the data mining communities~\cite{mandros2017discovering, reimherr}. The works in the database community study how to infer FDs that a dataset instance $D$ does not violate. These approaches are well-suited for database normalization purposes and for applications where strong closed-world assumptions on the given dataset $D$ hold. In contrast, the data mining community views FDs as statistical dependencies manifested in a dataset and has focused on information theoretic measures to estimate FDs. These approaches are more suited for data profiling and data cleaning applications. In this paper, we focus on FDs that correspond to statistical dependencies in the generating distribution of a given dataset. 

\paragraph{Challenges} Inferring FDs from data observations poses many challenges. First, to discover FDs one needs to identify an appropriate order of the attributes that captures the directionality of functional dependencies in a dataset. This leads to a computational complexity that scales exponentially in the number of attributes in a dataset. To address the exponential complexity of FD discovery, existing methods rely on pruning methods to search over the lattice of attribute combinations~\cite{kruse2018efficient, mandros2017discovering}. Despite the use of pruning many of the existing methods are shown to exhibit poor scalability as the number of columns increases~\cite{kruse2018efficient, mandros2017discovering}. 

Second, FDs capture deterministic relations between attributes. However, in real-world datasets missing or erroneous values introduce uncertainty to these relations. This poses a challenge as noise can lead to the discovery of spurious FDs or to low recall with respect to the true FDs in a dataset. To deal with missing values and erroneous data, existing FD discovery methods focus on identifying {\em approximate} FDs, i.e., dependencies that hold with high probability in a given dataset. To identify approximate FDs, existing methods either limit their search over clean subsets of the data~\cite{papenbrock2015functional} or employ a combination of sampling methods with error modeling~\cite{papenbrock2016hybrid, kruse2018efficient}. These methods are robust to noisy data. However, their performance, in terms of runtime and accuracy, is sensitive to factors such as sample sizes, prior assumptions on error rates, and the amount of records available in the input dataset. This makes these methods cumbersome to tune and apply to heterogeneous datasets with varying number of attributes, records, and errors.

Finally, most dependency measures used in FD discovery, such as co-occurrence counts~\cite{kruse2018efficient} or criteria based on mutual information~\cite{Cavallo:1987:TPD:645914.671645} promote complex dependency structures~\cite{mandros2017discovering}. The use of such measures leads to the discovery of spurious FDs in which the determinant set contains a large number of attributes. Such FDs are hard for humans to interpret and validate, especially when the goal is to use these FDs in downstream data preparation tasks. To avoid overfitting to complex FDs existing methods rely on post-processing procedures to simplify the structure of discovered FDs or ranking based solutions. The most common approach is to identify {\em minimal FDs}~\cite{papenbrock2015functional}. An FD $X \rightarrow Y$ is said to be minimal if no subset of $X$ determines $Y$. In many cases, this criterion is also integrated with search over the set of possible FDs for efficient pruning of the search space~\cite{papenbrock2016hybrid, kruse2018efficient}. Minimality is shown to be effective in practice, however, it does not guarantee that the overall set of discovered FDs will be parsimonious~\cite{kruse2018efficient}.

\paragraph{Our Contributions} We propose \system, a framework that relies on {\em structure learning}~\cite{Koller:2009:PGM:1795555} to solve FD discovery. Specifically, we leverage the strong dependencies that FDs introduce among attributes, introduce a probabilistic graphical model to capture these dependencies, and show that discovering FDs is equivalent to {\em learning the graph structure} of this model. {\em A key result in our work is to model the distribution that FDs impose over pairs of records instead of the joint distribution over the attribute-values of the input dataset}.

\system's model has {\em one binary random variable for each attribute} in the input dataset and expresses correlations amongst random variables via a graph that relates random variables in a linear way. We leverage linear dependencies to recover the directionality of FDs. Given a noisy dataset, \system proceeds in two steps: First, it estimates the {\em undirected form} of the graph that corresponds to the FD model of the input dataset. This is done by estimating the {\em inverse covariance matrix} of the joint distribution of the random variables that correspond to our FD model. Second, our FD discovery method finds a factorization of the inverse covariance matrix that imposes a sparse linear structure to the FD model, and thus, allows us to obtain parsimonious FDs.

We present an extensive experimental evaluation of \system. First, we compare our method against state-of-the-art methods from both the database and data mining literature over a diverse array of synthetic and real-world datasets with varying number of attributes, domain sizes, records, and amount of errors. We find that \system scales to large data instances with hundreds of attributes and yields an average $F_1$ improvement in discovering true FDs of more than $2\times$ compared to competing methods. 

We also examine the effectiveness of \system on downstream data preparation tasks. Specifically, we apply our FD discovery method on the task of weakly supervised data repairing. Recent work~\cite{hc} showed that integrity constraints (including functional dependencies) can be used to obtain noisy labeled data which can in turn be used to obtain state-of-the-art machine learning-based data repairing systems. We show that dependencies discovered via our method lead to high-quality repairs that are comparable to manually specified dependencies. This demonstrates that our FD discovery method offers a viable solution to automating weakly supervised data preparation tasks.

\paragraph{Outline} In Section~\ref{sec:prelims}, we discuss necessary background. In Section~\ref{sec:overview}, we formalize the problem of FD discovery and provide an overview of \system. In Section~\ref{sec:model}, we introduce the probabilistic model at the core of \system and the structure learning method we use to infer its graphical structure. Finally, in Section~\ref{sec:exps}, we present an experimental evaluation of \system, and conclude in Section~\ref{sec:conclusions}.
\section{Preliminaries}\label{sec:prelims}
We review some basic background material and introducing notation for the structure learning problem studied in this paper.

\subsection{Functional Dependencies}\label{sec:fds}

We review the concept of functional dependencies and related probabilistic interpretations. We consider a dataset $D$ that follows a relational schema $R$. An FD $\mathbf{X} \rightarrow Y$ is a statement over the set of attributes $\mathbf{X} \subseteq R$ and an attribute $Y \in R$ denoting that all tuples in $\mathbf{X}$ uniquely determine the values in $Y$~\cite{Garcia-Molina:1999:DSI:553977, papenbrock2015functional}. Formally, we consider $t_i[Y]$ to be the value of tuple $t_i \in D$ for attribute $Y$ ; the FD $X \rightarrow Y$ holds iff for all pairs of tuples $t_i, t_j \in D$ the following holds: if $\bigwedge_{A \in \mathbf{X}} t_i[A] = t_j[A]$ then $t_i[Y] = t_j[Y]$. A functional dependency $\mathbf{X} \rightarrow Y$ is {\em minimal} if no subset of $\mathbf{X}$ determines $Y$, and it is {\em non-trivial} if $Y \notin \mathbf{X}$. Under this logic-based interpretation, to discover all FDs in a dataset, it suffices to discover all minimal, non-trivial FDs. This interpretation makes strong closed-world assumptions and aims to find all FDs that hold in $D$. It does not aim to find FDs that hold in the generating distribution of $D$.

To relax these closed-world assumption, a probabilistic interpretation of FDs can be adopted. Let each attribute $A \in R$ have a domain $V(A)$ and the domain $V(\mathbf{X})$ of a set of attributes $\mathbf{X} = \{A_1, A_2, \dots, A_k\} \subseteq R$ be defined as $V(\mathbf{X}) = V(A_1) \times V(A_2) \times \dots \times V(A_k)$. Also, assume that every instance $D$ of $R$ is associated with a probability density $f_{R}(D)$ such that these densities form a valid probability distribution $P_R$. Given the distribution $P_R$, we say that an FD $\mathbf{X} \rightarrow Y$, with $\mathbf{X} \subseteq R$ and $Y \in R$, holds if there is a function $\phi: V(\mathbf{X}) \rightarrow V(Y)$ such that for all $\mathbf{x} \in V(\mathbf{X})$:
\begin{equation}\label{eq:old_fds}
P_R(Y=y | \mathbf{X}=\mathbf{x}) = 
\begin{cases}
1, \text{~when~} y = \phi(\mathbf{x})\\
0, \text{~otherwise}
\end{cases}
\end{equation}

This probabilistic definition represents a hard constraint that is not robust to noisy data. To relax this, a series of works have adopted information theoretic measures for FDs~\cite{Cavallo:1987:TPD:645914.671645, mandros2017discovering} by considering the ratio $F(\mathbf{X},Y) = \frac{H(Y) - H(Y|\mathbf{X})}{H(Y)}$ of the mutual information $H(Y) - H(Y|\mathbf{X})$ between $Y$ and $\mathbf{X}$ (where $H(Y|\mathbf{X})$ is the conditional entropy of $Y$ given $\mathbf{X}$) and the entropy $H(Y)$ of $Y$. To discover FDs one needs to identify sets of attributes $(\mathbf{X}, Y)$ in $R$ such that $F(\mathbf{X}, Y) = 1$. This requires estimating the entropy $H(Y)$ and conditional entropy $H(X|Y)$ from a given instance $D$ of $R$. We also adopt a probabilistic interpretation of FDs but build upon the framework of probabilistic graphical models to define FD discovery.

\subsection{Probabilistic Graphical Models}\label{sec:pgms}
We review key concepts in probabilistic graphical models~\cite{Koller:2009:PGM:1795555}.

\paragraph{Undirected Graphs} Let $P(x_1, \dots, x_m)$ be a probability distribution and $G = (V,E)$ an undirected graph where $V = \{1, \cdots, m\}$ and $E \subseteq V \times V$. We say that $G$ is a {\em conditional independence graph} for $P$ if: For all disjoint triples $(A,B,S) \subseteq V$ such that $S$ separates $A$ from $B$ in $G$ we have that $X_A$ and $X_B$ are independent given $X_S$, where $X_C = \{X_j : j \in C\}$ for any subset $C \subseteq V$. We also say that $G$ {\em represents} the distribution $P$. When $P$ is a strictly positive distribution (i.e., $P(x_1, x_2, \dots, x_m) > 0 $ for all $(x_1, \dots, x_m)$), then we have that $P(x_1, \dots, x_m) = \prod_{C \in \mathit{C}}\psi_{C}(x_C)$ for some potential functions $\{\psi_{C}: C \in \mathit{C}\}$ defined over the set of cliques $\mathit{X}$ of $G$. Undirected graphical models are also known as {\em Markov Random Fields}.

\paragraph{Directed Acyclic Graphs} We now consider a {\em directed} graph $G = (V,E)$. We say that $G$ is a {\em directed acyclic graph} (DAG) if there are no directed paths starting and ending at the same node. For each node $j \in V$ we define $\mathtt{Pa}(j) = \{k \in V: (k,j) \in E\}$ be the {\em parent set} of $j$, and write $\mathtt{Pa}_G(j)$ to emphasize the dependence on the structure of $G$. A DAG $G$ {\em represents} a distribution $P(x_1, \dots, x_m)$ if $P(x_1, \dots, x_m) \propto \prod_{j=1}^m P(x_j|x_{\mathtt{Pa}(j)})$. This factorization implies that given an observation for all parent nodes $X_{\mathtt{Pa}(j)}$ of $j$, $X_j$ is independent of all non-descendant nodes (i.e., nodes that cannot be reached via a directed path from $j$) excluding $\mathtt{Pa}(j)$.

\paragraph{Learning Parsimonious Graph Structures} Graphical models can encode simple or low-dimensional models. The complexity of a graphical model is related to the number of edges in $G$. It is easier to understand this notion of complexity if one considers the connection between graphical models and {\em generalized linear models} (GLIMs). An example of this connection is the Gaussian Markov Random Field model~\cite{Koller:2009:PGM:1795555, rue2005gaussian}. In GLIMs, parsimony is achieved by forcing the {\em inverse covariance matrix} (a.k.a. precision matrix) $\Theta = \Sigma^{-1}$ of the model to be sparse. This is because the conditional dependencies amongst the variables in the model are captured in the off-diagonal entries of the inverse covariance matrix $\Theta$. Zero off-diagonal entries in $\Theta$ represent conditional independencies amongst the variables of the model. Given this observation and the connection of Graphical Models to GLIMs, one can learn a parsimonious structure for a graphical model by obtaining a sparse estimate of the models inverse covariance matrix $\Theta$ from observed data. Many techniques have been proposed to obtain a sparse estimate for $\Theta$~\cite{pourahmadi2011} ranging from optimization methods~\cite{meinshausen2006high} to regression methods~\cite{friedman2008sparse}.
\section{The A\lowercase{uto}FD Framework}\label{sec:overview}
We formalize the problem of functional dependency discovery and provide an overview of \system. 

\subsection{Problem Statement}\label{sec:prob_statement}
We consider a relational schema $R$ associated with a probability distribution $P_R$. We assume access to a noisy dataset $D^{\prime}$ that follows schema $R$ and is generated by the following process: first a clean dataset $D$ is sampled from $P_R$ and a noisy channel model introduces noise in $D$ to generate obtain $D^{\prime}$. We assume that $D$ and $D^{\prime}$ have the same cells but cells in $D^{\prime}$ may have different values than their clean counterparts. We consider an {\em error} in $D^{\prime}$ to correspond to a cell $c$ for which $D^{\prime}(c) \neq D(c)$. This generative process is also considered in the database literature to model the creation of noisy datasets~\cite{icdt}. 

Given a noisy data instance $D^{\prime}$, our goal is to identify the functional dependencies that characterize the distribution $P_R$ that generated the clean version of $D$. In our work, we combine the probability-based and logic-based interpretations of FDs (see Section~\ref{sec:prelims}). For any pair of tuples $t_i$ and $t_j$ sampled from $P_R$, we denote $I_{ij} = \mathbbm{1}(t_i[Y] = t_j[Y])$ where $\mathbbm{1}(\cdot)$ is the indicator function, and denote $t_i[\mathbf{X}]$ the value assignment for attributes $\mathbf{X}$ in tuple $t_i$. We say that $t_i[\mathbf{X}] = t_j[\mathbf{X}]$ iff $\bigwedge_{A \in \mathbf{X}} t_i[A] = t_j[A]= \mathtt{True}$. Given a distribution $P_R$, we say that an FD $\mathbf{X} \rightarrow Y$, with $\mathbf{X} \subseteq R$ and $Y \in R$, holds for $P_R$ if for all pairs of tuples $t_i, t_j$ in $R$ we have that 

\begin{equation}
	\text{Pr}(I_{ij} = 1 ; t_i[\mathbf{X}], t_j[\mathbf{X}]) \propto 
\begin{cases}
1, \text{~when~} t_i[\mathbf{X}] = t_j[\mathbf{X}]\\
\theta, \text{~otherwise}
\end{cases}
\label{eq:condition}
\end{equation}
with $\theta = \sum_{y \in V(Y)}P_R(y; t_i[\mathbf{X}]) \cdot P_R(y; t_j[\mathbf{X}])$. This condition states that the two random events $\bigwedge_{A \in \mathbf{X}} t_i[A] = t_j[A]$ and $\mathbbm{1}(t_i[Y] = t_j[Y])$ are deterministically correlated when the FD $\mathbf{X} \rightarrow Y$ holds, otherwise they are independent. Under this interpretation, the problem of FD discovery corresponds to learning the structural dependencies amongst attributes of $R$ that satisfy the above condition. 

\begin{figure*}
\includegraphics[width=0.9\textwidth]{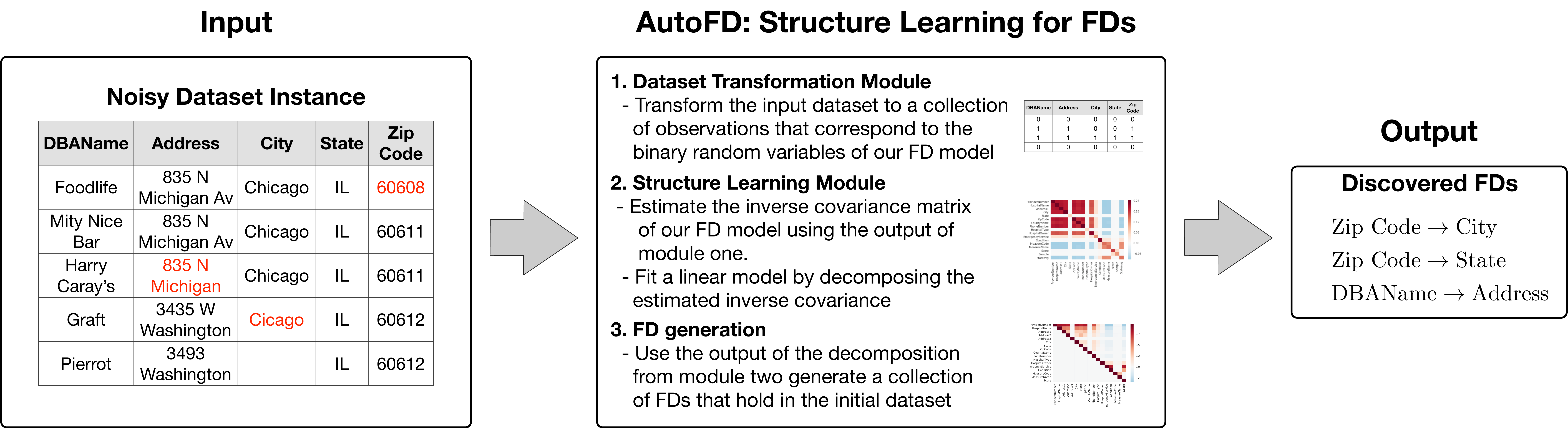}
\caption{An overview of our structure learning framework for FD discovery}
\label{fig:framework}
\end{figure*}

\subsection{Solution Overview}\label{sec:sol_overview}
We leverage the above probabilistic definition of FDs and build upon structure learning to solve FD discovery. An overview of our framework is shown in Figure~\ref{fig:framework}. The input to our framework is a noisy dataset and the output of our framework is a set of discovered FDs. The workflow of our framework follows three steps:

\vspace{2pt}\noindent\textbf{Dataset Transformation} First, we use the input dataset $D^{\prime}$ and generate a collection of samples that correspond to outcomes of the random events $\bigwedge_{A \in \mathbf{X}} t_i[A] = t_j[A] = \mathtt{True}$ and $t_i[Y] = t_j[Y]$. The output of this process is a new dataset $D_t$ that has one attribute for each attribute in $D^{\prime}$ but in contrast to $D^{\prime}$ it only contains binary values. We describe this step in Section~\ref{sec:prob_model}.

\vspace{2pt}\noindent\textbf{Structure Learning} Dataset $D_t$ contains samples from the distribution of events $\bigwedge_{A \in \mathbf{X}} t_i[A] = t_j[A] = \mathtt{True}$ and $t_i[Y] = t_j[Y]$. We consider a probabilistic graphical model $\mathit{M}$ associated with a graph $G$ that represents these events (see Section~\ref{sec:prob_model}) and use the samples in $D_t$ to learn the structure of $G$. Here, we leverage the fact that that our model $\mathit{M}$ corresponds to a generalized linear model, and learn its structure by obtaining a sparse estimate of its inverse covariance matrix. We describe our structure learning method in Section~\ref{sec:strl}.

\vspace{2pt}\noindent\textbf{FD generation} Finally, we use the estimated inverse covariance matrix to generate a collection of FDs. We do so by considering the non-zero off-diagonal entries of the estimated inverse covariance matrix. The final output of our model is a collection of discovered FDs of the form $\mathbf{X} \rightarrow Y$ where $\mathbf{X} \subseteq R$ and $Y \in R$.

\section{FD Discovery in A\lowercase{uto}FD}\label{sec:model}
We first introduce the probabilistic graphical model that \system uses to represent FDs and then describe our approach to learning its structure. Finally, we discuss how our approach compares to a naive application of structure learning to FD discovery.

\subsection{The \system Model}\label{sec:prob_model}
\system's probabilistic graphical model is inspired by the FD definition described in Equation~\ref{eq:condition} and aims to capture the distribution of the random events $\bigwedge_{A \in \mathbf{X}} t_i[A] = t_j[A]$ and $\mathbbm{1}(t_i[Y] = t_j[Y])$. \system's model consists of random variables that model these two random events. The edges in the model represent statistical dependencies that capture the relation in Equation~\ref{eq:condition}.

We have one random variable per attribute in $R$. For each attribute $A \in R$, we denote $Z_{A} \in \{0,1\}$ the random event of sampling two tuples from distribution $P_R$ such that they have the same value for attribute $A$. In other words, for any sample $(t_i, t_j)$ from $P_R$, the {\em binary random variable} $Z_{A}$ takes $Z_{A} = 1$ iff $t_i[A] = t_j[A]$. We now define the edges over the set of binary random variables $\bigcup_{A \in R}\{Z_A\}$. Assume that the FD $F: \mathbf{X} \rightarrow Y$ holds and hence the correlation defined in Equation~\ref{eq:condition} holds. We represent the dependency between attributes $\mathbf{X}$ and $Y$ be having a directed edge from each attribute $A \in \mathbf{X}$ to attribute $Y$. Each true FD in the data generating distribution corresponds to a directed subgraph with V-structure. Let $\mathbf{X} = \{X_1, X_2, \dots, X_k\}$. For Equation~\ref{eq:condition} to hold, the entries of the conditional probability table $\Pi_F$ for the subgraph corresponding to FD $F$ should such that: $\Pi_F(Z_Y = 1, Z_{X_1} = 1, Z_{X_2 = 1}, \dots, Z_{X_k}=1) = 1$, $\Pi_F(Z_Y = 0, Z_{X_1} = 1, Z_{X_2 = 1}, \dots, Z_{X_k}=1) = 0$, and all other entries should be set such that they force an independence structure. We assume {\em acyclic FDs}, i.e., we do not allow for sets of FDs such as $A \rightarrow B$ and $B,C \rightarrow A$. As a result, the graphical structure of this model corresponds to a directed probabilistic graphical model where each FD introduces a V-structure subgraph. We assume a global order over the FDs which also defines the global order of the random variables in the above model.

Our goal is to learn the graphical structure of the model described above. However, learning the structure of a directed graphical model with V-structure patterns is NP-hard~\cite{Chickering:2004:LLB:1005332.1044703}. In fact, it is only for tree-based directed graphical models that one can obtain guarantees for graph-based structure learning methods~\cite{Koller:2009:PGM:1795555}. Given this hardness result, we turn our attention to structure learning for parsimonious generalized linear models (see Section~\ref{sec:prelims}. Specifically, we relax our initial model to a {\em linear structural equation model} that approximates the condition in Equation~\ref{eq:condition}. This is the actual model that \system uses for FD discovery. We next describe this relaxed model.

First, we relax the random variables $\{ Z_{A} \}_{A \in R}$ to take values in $[0,1]$ instead of $\{0,1\}$. Second, we have that when $\bigwedge_{A \in \mathbf{X}} t_i[A] = t_j[A] = \bigwedge_{A \in \mathbf{X}} Z_{A} = \mathtt{True}$ it must be that $\mathbbm{1}(t_i[Y] = t_j[Y]) = Z_{Y} = 1$. To represent this condition for real-values random variables we rely on {\em soft logic}~\cite{Bach:2017:HMR:3122009.3176853}. Soft logic allows continuous truth values from the interval $[0, 1]$ instead of ${0, 1}$, and the Boolean logic operators are reformulated as: $A \wedge B = \max\{A+ B -1, 0\}$,~$A \lor B = \min\{A + B, 1\}$, $~A_1 \wedge A_2 \wedge \dots A_k = \frac{1}{k}\sum_{i} A_i$, and $\neg A = 1 - A$. Based on this formulation of conjunction, we can approximate the condition in Equation~\ref{eq:condition} by requiring that $Z_Y = \frac{1}{|\mathbf{X}|}\sum_{A \in \mathbf{X}}Z_{A}$ when the FD $\mathbf{X} \rightarrow Y$ holds. We leverage this relaxed condition to derive \system's model for FD discovery.

We consider the random vector $\mathbf{Z} = \{Z_{A_1}, Z_{A_2}, \dots, Z_{A_{|R|}}\} \in [0,1]^l$ that corresponds to the random variables associated with the attributes in schema $R$. Based on the aforementioned relaxed condition, FDs force this random vector to follow a {\em linear structured equation model}. Hence, we can write that: 

\begin{equation}\label{eq:lmodel}
	\mathbf{Z} = B^T\mathbf{Z} + \epsilon,
\end{equation}
where we assume that $E[\epsilon] = 0$ and $\epsilon_j \ci (Z_{A_1}, \dots, Z_{A_{j-1}})$ for all $j$, where $\ci$ denotes conditional independence. Since our model corresponds to a directed graphical model, matrix $B$ is a strictly upper triangular matrix. $B$ is known as the {\em autoregression matrix} of the system~\cite{loh2014high}. For DAG $G$ with vertex set $V = \{Z_{A_1}, Z_{A_2}, \dots, Z_{A_{|R|}}\}$ and edge set $E = \{(j,k) : B_{jk} \neq 0\}$, the joint distribution factorizes as $ P(Z_{A_1}, \dots, Z_{A_{|R|}}) = \prod_{j=1}^{|R|}P(Z_{A_j}|Z_{A_1},\dots, Z_{A_{j-1}})$. Given samples $\{\mathbf{Z}^i\}_{i=1}^{N}$, our goal is to infer the unknown matrix $B$.
 
\subsection{Structure Learning in \system}\label{sec:strl}

Our structure learning algorithm follows from results in statistical learning theory. We build upon a recent result of Loh and Buehlmann~\cite{loh2014high} on learning the structure of linear causal networks via inverse covariance estimation. Given a linear model as the one shown in Equation~\ref{eq:lmodel}, it can be shown that the inverse covariance matrix $\Theta = \Sigma^{-1}$ of the model can be written as:

\begin{equation}\label{eq:decomp}
	\Theta = \Sigma^{-1} = (I - B)\Omega^{-1}(I-B)^{T}
\end{equation}

where $I$ is the identity matrix, $B$ is the autoregression matrix of the model, and $\Omega = \text{cov}[\epsilon]$ with $\text{cov}[\cdot]$ denoting the covariance matrix. This decomposition of $\Theta$ is also commonly used in generalized linear models for learning parsimonious models~\cite{pourahmadi2011}. 

Given Equation~\ref{eq:decomp}, FD discovery in \system proceeds as follows: First, we transform the sample data records in the input dataset $D^{\prime}$ to samples $\{\mathbf{Z}^i\}_{i=1}^{N}$ for the linear model in Equation~\ref{eq:lmodel} (see Algorithm~\ref{alg:transform}); Second, we obtain an estimate $\hat{\Theta}$ of the inverse covariance matrix and factorize the estimate $\hat{\Theta}$ to obtain an estimate of the autoregression matrix $\hat{B}$; Third, we use the estimated matrix $\hat{B}$ to generate FDs (see Algorithm~\ref{alg:generation}).

\begin{algorithm}
\small
\SetAlgoLined
\KwIn{A noisy relational dataset $D^{\prime}$ following schema $R$.}
\KwOut{A set of FDs of the form $\mathbf{X} \rightarrow Y$ on $R$.}
Set $D_t \leftarrow \texttt{Transform}(D^{\prime})$ (See Alg. 2)\;
Obtain an estimate $\hat{\Theta}$ of the inverse covariance matrix (e.g., using Graphical Lasso)\;
Factorize $\hat{\Theta} = UDU^{T}$ with $U$ being upper triangular\;
Set $\hat{B} = I - U$\;
Set $\text{Discovered FDs} \leftarrow \texttt{GenerateFDs}(B)$ (See Alg. 3)\;
\textbf{return} Discovered FDs
\caption{FD discovery with \system}
\label{alg:framework}
\end{algorithm}

An overview of \system's FD discovery method is shown in Algorithm~\ref{alg:framework}. The structure learning part in this algorithm proceeds as follows: Suppose we have $N$ observations and let $S$ by the empirical covariance matrix of these observations. It is a standard result~\cite{meinshausen2006high} that the sparse inverse covariance $\theta$ can be estimated by solving the following optimization problem: $\min_{\Theta \succ  0} f(\Theta):=-\log\det(\Theta) + tr(S\Theta) + \lambda\left \| \Theta \right \|_1$. Friendman et al.~\cite{friedman2008sparse} have shown that one can approximate the solution to this problem by solving a series of LASSO problems. This method is known as {\em Graphical Lasso} and is one of the de-facto algorithms for structure learning. Graphical Lasso is shown to scale favorably to large instances and hence is appropriate for our setting. In our experimental evaluation, we show that our methods can scale to datasets with millions of records and tens of attributes. Given the estimated inverse covariance matrix $\hat{\Theta}$, we use the Bunch-Kaufman algorithm to obtain a factorization of $\hat{\Theta}$ and obtain an estimate for the autoregression matrix $\hat{B}$. To generate FDs from $\hat{B}$ we use Algorithm~\ref{alg:generation}. 

\begin{algorithm}
\small
\SetAlgoLined
\KwIn{A dataset $D$ with $n$ rows and $k$ columns}
\KwOut{A dataset $D_t$ with $n \cdot k$ rows and $k$ columns}

$A \leftarrow $ columns $[A_1,...,A_k]$\;
$D \leftarrow$ shuffle rows of $D$\;
$D_t \leftarrow \emptyset$\;
\For{i = 1 : k}{
	$D_i \leftarrow$ sort $D$ by attribute $A_i$\;
	$D_{i\_shift} \leftarrow $ circular shift of rows in $D_i$ by 1\;
	\For{j = 1 : n}{
	    \For{l = 1 :k}{
	         $D_t[(i-1)\cdot n + j, l] \leftarrow \mathbbm{1}\left(D_i[j,l] = D_{i\_shift}[j,l]\right)$\;
	    }
 	}
 }
 \KwRet{$D_t$}
\caption{Data Transformation}
\label{alg:transform}
\end{algorithm}

We now turn our attention to how we transform the input dataset $D^{\prime}$ into a collection $D_t$ of observations for the linear model of \system (see Algorithm~\ref{alg:transform}). We use the differences of pairs of tuples in dataset $D^{\prime}$ to generate $D_t$. As shown in Algorithm~\ref{alg:transform}, we perform a self-join over the input dataset and consider the value differences between the generated pairs of tuples to obtain observations for the random variables $\mathbf{Z}$ in \system's probabilistic model. Our method can support diverse data types (e.g., categorical, real-values, text data, binary data, or mixtures of those) as we can use a different difference operation for each of these types.

\begin{algorithm}
\small
\SetAlgoLined
\KwIn{An autoregression matrix $B$ of dimensions $n \times m$, A schema $R$}
\KwOut{A collection of FDs}
$\texttt{FDs} \leftarrow \emptyset$\;
\For{j = 1 : m}{
    Set the column vector $b_j \leftarrow (B_{1,j}, B_{2,j}, \dots, B_{j-1,j})$ \;
    $\mathbf{X} \leftarrow $ Take the attributes in $R$ that corresponds to non-zero entries in $b_j$\;
    Let $A_j$ be the attribute in $R$ with coordinate $j$ \;
    \If{$\mathbf{X} \neq \emptyset$}{
         $\texttt{FDs} \leftarrow \texttt{FDs} \cup \{\mathbf{X} \rightarrow A_j\}$\;
    }
 }
 \KwRet{$\texttt{FDs}$}
\caption{FD generation}
\label{alg:generation}
\end{algorithm}

\subsection{Discussion}\label{sec:discussion}
There are certain benefits that \system's model offers when compared to applying structure learning directly on $D^{\prime}$.

Our transformation allows us to solve a structure learning where we have access to an increased amount of training data. As we will show in Section~\ref{sec:exps}, existing methods are not robust when the sample size is small. Information-theoretic approaches, such as the one by Mandros et al.~\cite{mandros2017discovering}, tend to assign a low-confidence score to FDs for small sample sizes. Hence, they exhibit limited recall. 

Structure learning for the model described in Section~\ref{sec:prob_model} enjoys better sample complexity than applying structure learning on the raw input dataset. We focus on the case of discrete random variables to explain this argument. Let $k$ be the size of the domain of the variables. The sample complexity of state-of-the-art structure learning algorithms is proportional to $k^4$~\cite{wu2018sparse}. Our model restricts the domain of the random variables to be $k=2$, and hence, yields better sample complexity than applying structure learning directly on the raw input. We demonstrate this experimentally in Section~\ref{sec:exps}.
\section{Experiments}\label{sec:exps}
We compare \system against several FD discovery methods on diverse datasets. The main points we seek to validate are: (1) does structure learning enables us to discover FDs with accurately (i.e., with high precision and recall), (2) can \system scale to large datasets, and (3) can \system provide FDs that are useful for downstream data preparation tasks. We also perform  micro-benchmark experiments to examine the effectiveness and sensitivity of our model. 

\subsection{Experimental Setup}\label{exp_setup}
\textbf{Datasets}: We use both synthetic and real-world datasets in our experiments. Our synthetic datasets aim to capture different data properties with respect to four key factors that affect the performance of FD discovery algorithms: (1) Noise Rate (denoted by $n$). It stresses the robustness of FD discovery methods; (2) Number of Tuples (denoted by $t$). It affects the sample size available to the FD discovery methods; (3) Number of Attributes (denoted by $r$); It stresses the scalability of FD discovery methods; (4) Domain Cardinality (denoted by $d$) of the left-hand side $\mathbf{X}$ for an FD; It evaluates the sample complexity of FD methods. For our end-to-end evaluation (see Section~\ref{sec:end2end}), we consider 24 different setting combinations for these four dimensions (summarized in Table~\ref{tab:synthetic_settings}). For each setting we use a mixture of FDs $\mathbf{X} \rightarrow Y$ for which the cardinality of $\mathbf{X}$ ranges from one to three.

\begin{table}
  \small
  \caption{The different settings we consider for synthetic datasets. We use the description in parenthesis to denote each of these settings in our experiments.}
  \label{tab:synthetic_settings}
  \begin{tabular}{l l}
    \toprule
    Property & Settings \\
    \midrule
    Noise Rate (n) & 0\% (Zero), 1\% (Low), 30\% (High) \\
    Tuples (t) & 1,000 (Small), 100,000 (Large) \\
    Attributes (r) & 8-16 (Small), 40-80 (Large) \\
    Domain Cardinality for FD (d) & 64-216 (Small), 1,000-1,728 (Large)\\
  \bottomrule
\end{tabular}
\end{table}

We follow the next process to generate synthetic data. Given a schema with $r$ attributes our generator first assigns a global order to these attributes and splits the ordered attributes in consecutive attribute sets, whose size is between two and four (so that we obey the cardinality of the FD as we discussed above). Let $(\mathbf{X},Y)$ be the attributes in such a split. Our generator samples a value $v$ from the range associated with the setting for Domain Cardinality and assigns a domain to each attribute in $\mathbf{X}$ such that the cartesian product of the attribute values corresponds to that value. It also assigns the domain size of $Y$ to be $v$. 

To simulate real-world data, we introduce FD dependencies as well as correlations in the splits obtained by the above process. For half of the ($\mathbf{X}$, $Y$) groups generated via the above process, we introduce FD-based dependencies that satisfy the property in Equation~\ref{eq:old_fds}. We do so by assigning each value $l \in \text{dom}(\mathbf{X})$ to a value $r_0 \in \text{dom}(Y)$ uniformly at random and generating $t$ samples, where $t$ is the value for the Tuples parameter. For the remainder of those groups we force the following conditional probability distribution: We assign each value $l \in \text{dom}(\mathbf{X})$ to a value $r_0 \in \text{dom}(Y)$. Then we generate $t$ samples with $P(Y=r_0~|~\mathbf{X}=l) =  \rho$ and $P(Y\neq r_0~|~\mathbf{X}=l) = \frac{1-\rho}{|\text{dom}(Y) - 1|}$. Here, $\rho$ is a hyper-parameter that is sampled uniformly at random from $[0,0.85]$. This process allows us to mix FDs with other correlations, and hence, evaluate the ability of FD discovery mechanisms to differentiate between true FDs and strong correlations. Finally, to test how robust FD discovery algorithms are to noise, we randomly flip cells that correspond to attributes that participate in true FDs to a different value from their domain. The percentage of flipped cells is controlled by the Noise Rate setting.

\begin{table}
  \small
  \caption{Real-world datasets for our experiments.}
  \label{tab:real_data}
  \begin{tabular}{cccc}
    \toprule
    Dataset & Size & Attributes & Errors (\# of cells)\\
    \midrule
    Hospital & 1,000 & 19 & 504\\
    Food & 170,945 & 15 & 31,296\\
    Physician & 2,071,849 & 18 & 174,557\\
  \bottomrule
  \end{tabular}
\end{table}

For real-world datasets, we use three noisy datasets. Table~\ref{tab:synthetic_settings} provides information for these datasets.  (1) The Hospital dataset is a small benchmark dataset used in several data cleaning papers~\cite{hc,holistic}. Errors are artificially introduced by injecting typos; (2) The Food dataset contains information on food establishments in Chicago. Errors correspond to typos; (3) The Physician dataset form Medicare.gov\footnote{https://data.medicare.gov/data/physician-compare}. Errors correspond to typos and null values.

\noindent\textbf{Methods}: We compare \system against:

\vspace{2pt}\noindent\textbf{PYRO}~\cite{kruse2018efficient}: PYRO is the state-of-the-art FD discovery method in the database community~\cite{kruse2018efficient}. The code we used for experiments is released by the authors.\footnote{https://github.com/HPI-Information-Systems/pyro/releases}. The scalability of the algorithm is controlled via an error rate hyper-parameter.

\vspace{2pt}\noindent\textbf{Reliable Fraction of Information (RFI)}\cite{mandros2017discovering}: This method is the state-of-the-art FD discovery approach in the data mining community. It relies on an information theoretic score to identify FDs and uses an approximation scheme to optimize performance. The approximation ratio is controlled by a user specified hyper-parameter $\alpha$. We evaluate RFI for $\alpha \in \{0.3, 0.5, 1\}$ where a value of $1.0$ corresponds to no approximation. The code we used is released by the authors.\footnote{http://eda.mmci.uni-saarland.de/prj/dora/} This implementation discovers FDs for one attribute at a time. To discover all FDs in a dataset, we run the provided method once per attribute.

\vspace{2pt}\noindent\textbf{Graphical Lasso (GL)}: We also evaluate a state-of-the-art structure learning algorithm on the raw input dataset $D^\prime$. Graphical Lasso provides as with an estimate of the inverse covariance $\hat{\Theta}$ of that problem. Graphical Lasso is shown to recover the true structure of the undirected graphical model that represents the distribution that corresponds to $D^\prime$~\cite{wu2018sparse}. In this case we cannot factorize $\hat{\Theta}$ to generate FDs. To find FDs that determine attribute $Y$, we take the neighborhood (as defined by $\hat{\Theta}$ of the corresponding random variable and perform a local graph search to find high-score directed structures~\cite{Koller:2009:PGM:1795555}.

\noindent\textbf{Evaluation Setup}: To measure accuracy, we use Precision (P) defined as the fraction of correctly discovered FDs by the total number of discovered FDs; Recall (R) defined as the fraction of correctly discovered FDs by the total number of true FDs in the dataset; and $F_1$ is defined as $2PR/(P+R)$. For synthetic dataset, each setting has five corresponding dataset instances. To ensure that we maintain the coupling amongst Precision, Recall, and $F_1$, we report the median performance. For all methods, we fine-tuned their hyper-parameters to optimize performance. In the case of Pyro we consulted the authors for this process. All experiments without specific description were executed on a machine with Two Intel Xeon Silver 4114 10-core CPUs at 2.20 GHz and 192GB Memory. Every time we run 2 datasets in parallel and each dataset is assigned 16 isolated threads and 93GB Memory. 


\subsection{End-to-end Performance}\label{sec:end2end}
We evaluate the performance of \system against competing approaches on the synthetic and real-world data described above. We first present quantitative results on the synthetic data (since we know the exact FDs) and then present qualitative results on the real-world datasets.

\begin{table}[t]
\center
\scriptsize
\caption{Precision, Recall and $F_1$-score of different methods for different synthetic settings. A description of the different settings is provided in Table~\ref{tab:synthetic_settings}.}
\label{tab:exp2}
\begin{threeparttable}
\begin{tabular}{c | c | c | c |c c c| c c c c}
n & t & r & d & & \system & GL & PYRO & \shortstack{RFI \\($\alpha = 0.3$)} & \shortstack{RFI \\($\alpha = 0.5$)} & \shortstack{RFI \\($\alpha = 1.0$)} \\ \hline \hline
\multirow{24}{*}{\rotatebox[origin=c]{90}{High}} & 
 \multirow{12}{*}{l} & 
 \multirow{6}{*}{l} & 
 \multirow{3}{*}{l} & 
 P &
 \bf 0.500 &  0.143 &  0.001 &  - &  - &  - \\
 & & & & R &
 \bf 1.000 &  0.100 &  0.200 &  - &  - &  - \\
 & & & & $F_1$ &
 \bf 0.667 &  0.118 &  0.001 &  - &  - &  - \\
\cline{4-11}
 & & & \multirow{3}{*}{s} & 
 P &
 \bf 0.435 &  0.353 &  0.001 &  - &  - &  - \\
 & & & & R &
 \bf 1.000 &  0.600 &  0.300 &  - &  - &  - \\
 & & & & $F_1$ &
 \bf 0.606 &  0.444 &  0.002 &  - &  - &  - \\
\cline{3-11}
 & & \multirow{6}{*}{s} & 
 \multirow{3}{*}{l} & 
 P &
 \bf 0.400 &  0.000 &  0.005 &  - &  - &  - \\
 & & & & R &
 \bf 0.500 &  0.000 &  0.250 &  - &  - &  - \\
 & & & & $F_1$ &
 \bf 0.500 &  0.000 &  0.009 &  - &  - &  - \\
\cline{4-11}
 & & & \multirow{3}{*}{s} & 
 P &
 \bf 0.500 &  0.333 &  0.006 &  - &  - &  - \\
 & & & & R &
 \bf 0.500 &  0.500 &  0.500 &  - &  - &  - \\
 & & & & $F_1$ &
 \bf 0.500 &  0.400 &  0.013 &  - &  - &  - \\
\cline{2-11}
 & \multirow{12}{*}{s} & 
 \multirow{6}{*}{l} & 
 \multirow{3}{*}{l} & 
 P &
 \bf 0.600 &  0.000 &  0.001 &  - &  - &  - \\
 & & & & R &
 \bf 0.400 &  0.000 &  0.400 &  - &  - &  - \\
 & & & & $F_1$ &
 \bf 0.471 &  0.000 &  0.002 &  - &  - &  - \\
\cline{4-11}
 & & & \multirow{3}{*}{s} & 
 P &
 \bf 0.304 &  0.000 &  0.001 &  - &  - &  - \\
 & & & & R &
 \bf 0.700 &  0.000 &  0.200 &  - &  - &  - \\
 & & & & $F_1$ &
 \bf 0.424 &  0.000 &  0.001 &  - &  - &  - \\
\cline{3-11}
 & & \multirow{6}{*}{s} & 
 \multirow{3}{*}{l} & 
 P &
 \bf 0.250 &  0.000 &  0.000 &  0.000 &  0.000 &  0.000 \\
 & & & & R &
 \bf 0.500 &  0.000 &  0.000 &  0.000 &  0.000 &  0.000 \\
 & & & & $F_1$ &
 \bf 0.333 &  0.000 &  0.000 &  0.000 &  0.000 &  0.000 \\
\cline{4-11}
 & & & \multirow{3}{*}{s} & 
 P &
 \bf 0.400 &  0.000 &  0.000 &  0.000 &  0.000 &  0.000 \\
 & & & & R &
 \bf 1.000 &  0.000 &  0.000 &  0.000 &  0.000 &  0.000 \\
 & & & & $F_1$ &
 \bf 0.571 &  0.000 &  0.000 &  0.000 &  0.000 &  0.000 \\
\cline{1-11}
\multirow{24}{*}{\rotatebox[origin=c]{90}{Low}} & 
 \multirow{12}{*}{l} & 
 \multirow{6}{*}{l} & 
 \multirow{3}{*}{l} & 
 P &
 \bf 0.400 &  0.364 &  0.000 &  - &  - &  - \\
 & & & & R &
 \bf 1.000 &  0.400 &  0.200 &  - &  - &  - \\
 & & & & $F_1$ &
 \bf 0.571 &  0.381 &  0.000 &  - &  - &  - \\
\cline{4-11}
 & & & \multirow{3}{*}{s} & 
 P &
 \bf 0.714 &  0.353 &  0.000 &  - &  - &  - \\
 & & & & R &
 \bf 1.000 &  0.600 &  1.000 &  - &  - &  - \\
 & & & & $F_1$ &
 \bf 0.833 &  0.444 &  0.000 &  - &  - &  - \\
\cline{3-11}
 & & \multirow{6}{*}{s} & 
 \multirow{3}{*}{l} & 
 P &
 \bf 0.667 &  0.333 &  0.008 &  0.375 &  - &  - \\
 & & & & R &
 \bf 1.000 &  0.500 &  0.500 &  0.750 &  - &  - \\
 & & & & $F_1$ &
 \bf 0.800 &  0.400 &  0.016 &  0.500 &  - &  - \\
\cline{4-11}
 & & & \multirow{3}{*}{s} & 
 P &
 \bf 1.000 &  0.500 &  0.002 &  1.000 &  - &  - \\
 & & & & R &
 0.500 &  \bf 1.000 &  1.000 &  1.000 &  - &  - \\
 & & & & $F_1$ &
 0.667 &  0.667 &  0.004 &  \bf 1.000 &  - &  - \\
\cline{2-11}
 & \multirow{12}{*}{s} & 
 \multirow{6}{*}{l} & 
 \multirow{3}{*}{l} & 
 P &
 \bf 0.533 &  0.017 &  0.000 &  - &  - &  - \\
 & & & & R &
 \bf 0.700 &  0.100 &  0.300 &  - &  - &  - \\
 & & & & $F_1$ &
 \bf 0.640 &  0.029 &  0.000 &  - &  - &  - \\
\cline{4-11}
 & & & \multirow{3}{*}{s} & 
 P &
 \bf 0.909 &  0.167 &  0.000 &  - &  - &  - \\
 & & & & R &
 \bf 1.000 &  0.100 &  1.000 &  - &  - &  - \\
 & & & & $F_1$ &
 \bf 0.952 &  0.143 &  0.000 &  - &  - &  - \\
\cline{3-11}
 & & \multirow{6}{*}{s} & 
 \multirow{3}{*}{l} & 
 P &
 \bf 0.667 &  0.000 &  0.008 &  0.250 &  0.250 &  0.250 \\
 & & & & R &
 \bf 1.000 &  0.000 &  0.500 &  0.500 &  0.500 &  0.500 \\
 & & & & $F_1$ &
 \bf 0.800 &  0.000 &  0.016 &  0.333 &  0.333 &  0.333 \\
\cline{4-11}
 & & & \multirow{3}{*}{s} & 
 P &
 \bf 1.000 &  0.000 &  0.005 &  0.143 &  0.286 &  0.286 \\
 & & & & R &
 \bf 1.000 &  0.000 &  1.000 &  0.500 &  1.000 &  1.000 \\
 & & & & $F_1$ &
 \bf 1.000 &  0.000 &  0.010 &  0.222 &  0.444 &  0.444 \\
\cline{1-11}
\multirow{24}{*}{\rotatebox[origin=c]{90}{Zero}} & 
 \multirow{12}{*}{l} & 
 \multirow{6}{*}{l} & 
 \multirow{3}{*}{l} & 
 P &
 \bf 0.667 &  0.214 &  - &  - &  - &  - \\
 & & & & R &
 \bf 0.600 &  0.300 &  - &  - &  - &  - \\
 & & & & $F_1$ &
 \bf 0.632 &  0.250 &  - &  - &  - &  - \\
\cline{4-11}
 & & & \multirow{3}{*}{s} & 
 P &
 \bf 0.667 &  0.421 &  - &  - &  - &  - \\
 & & & & R &
 \bf 1.000 &  0.800 &  - &  - &  - &  - \\
 & & & & $F_1$ &
 \bf 0.800 &  0.552 &  - &  - &  - &  - \\
\cline{3-11}
 & & \multirow{6}{*}{s} & 
 \multirow{3}{*}{l} & 
 P &
 \bf 1.000 &  0.667 &  0.000 &  - &  - &  - \\
 & & & & R &
 \bf 1.000 &  0.500 &  0.000 &  - &  - &  - \\
 & & & & $F_1$ &
 \bf 1.000 &  0.667 &  0.000 &  - &  - &  - \\
\cline{4-11}
 & & & \multirow{3}{*}{s} & 
 P &
 \bf 1.000 &  0.400 &  0.006 &  1.000 &  1.000 &  - \\
 & & & & R &
 \bf 1.000 &  1.000 &  0.500 &  0.500 &  0.500 &  - \\
 & & & & $F_1$ &
 \bf 1.000 &  0.500 &  0.012 &  0.667 &  0.667 &  - \\
\cline{2-11}
 & \multirow{12}{*}{s} & 
 \multirow{6}{*}{l} & 
 \multirow{3}{*}{l} & 
 P &
 \bf 0.714 &  0.017 &  0.000 &  - &  - &  - \\
 & & & & R &
 \bf 0.500 &  0.100 &  0.200 &  - &  - &  - \\
 & & & & $F_1$ &
 \bf 0.588 &  0.029 &  0.000 &  - &  - &  - \\
\cline{4-11}
 & & & \multirow{3}{*}{s} & 
 P &
 \bf 0.769 &  0.143 &  - &  - &  - &  - \\
 & & & & R &
 \bf 1.000 &  0.100 &  - &  - &  - &  - \\
 & & & & $F_1$ &
 \bf 0.870 &  0.118 &  - &  - &  - &  - \\
\cline{3-11}
 & & \multirow{6}{*}{s} & 
 \multirow{3}{*}{l} & 
 P &
 \bf 0.667 &  0.000 &  0.001 &  0.000 &  0.000 &  0.000 \\
 & & & & R &
 \bf 1.000 &  0.000 &  0.500 &  0.000 &  0.000 &  0.000 \\
 & & & & $F_1$ &
 \bf 0.800 &  0.000 &  0.003 &  0.000 &  0.000 &  0.000 \\
\cline{4-11}
 & & & \multirow{3}{*}{s} & 
 P &
 \bf 1.000 &  0.100 &  0.001 &  0.200 &  0.200 &  - \\
 & & & & R &
 \bf 1.000 &  0.500 &  0.500 &  0.500 &  0.500 &  - \\
 & & & & $F_1$ &
 \bf 1.000 &  0.167 &  0.003 &  0.286 &  0.286 &  - \\
\cline{1-11}
\end{tabular}
\begin{tablenotes}
  \item '-': method exceeds runtime limit (8 hours),  or runs out of memory, or output is more than 7 GB.
  \end{tablenotes}
\end{threeparttable}
\end{table}

\subsubsection{Accuracy}\label{exp2-1}
Table~\ref{tab:exp2} shows the precision, recall, and $F_1$-score obtained by different methods. As shown, \system consistently outperforms all other methods in terms of $F_1$-score across all settings, with an $F_1$ improvement of more than 2X on average. More importantly, we find that \system is less affected by limited sample sizes and high-cardinality domains compared to other FD discovery methods. In detail, we find that \system maintains good precision and recall for datasets with low amount of noises ($\leq$ 1\%) with an average precision of 85.52\% and an average recall of 99.75\%. Despite the fact that it exhibits an average $F_1$ drop of 27.38\% for datasets with high noise rate, \system still yields better precision and recall than competing methods. {\em This verifies our hypothesis that structure learning along with the data transformation step introduced in Section~\ref{sec:prob_model} leads to more a accurate FD discovery solution}.

We focus on the results for competing methods. We start with PYRO. To optimize PYRO's performance we set its error rate hyper-parameter to the noise level for each dataset. For low noise-rates PYRO may not terminate. We see that in most cases PYRO obtains high recall but low precision. This behavior is expected as PYRO follows a logic-based interpretation of FDs (see Section~\ref{sec:prelims}) and aims to discover all FDs that hold for a given dataset instance. It is not designed to find the true FDs in the data generating distribution or interpretable FDs for data preparation tasks. For example, for datasets with small number of attributes (8-16), PYRO finds 446 FDs on average, excluding the outputs ranging from 7.8 GB to 10 GB that we cannot handle, which may affect the performance in downstream data preparation tasks.

We now turn our attention to RFI. As shown, RFI exhibits poor scalability as in many cases it fails to terminate within 8 hours and in others it raises out-of-memory issues. For the cases that RFI terminates we find that it exhibits high precision for small cardinality domains when a large number of samples is available and the noise rate is low. As the sample size decreases or the noise rate increases we find that the performance of RFI drops significantly. We further investigated the performance of RFI for partial executions. Recall that due to the implementation of RFI, we have to run it for each attribute separately. We evaluated RFI's accuracy for each of the attributes processed within the 8-hour time window. Our findings are consistent with the aforementioned observation. The precision of RFI is very high but its recall is lower than \system. The main takeaway is that RFI has high sample complexity.

Finally, we see that the high sample complexity of structure learning on the raw input (see Section~\ref{sec:discussion}) leads to GL exhibiting low accuracy. This becomes more clear, if we compare the performance of GL with a large number of tuples to that with a small number of tuples while keeping other variables constant. We can see a consistent drop of performance when the data sample becomes limited. \emph{This validates our modeling choices for \system}.

\begin{table}[t]
\center
\scriptsize
\caption{Average runtime (in seconds) of different methods for different synthetic settings.}
\label{tab:synruntime}
\begin{threeparttable}
\begin{tabular}{c | c | c | c | c c | c c c c}
n & t & r & d & \system & GL & PYRO & \shortstack{RFI \\($\alpha = 0.3$)} & \shortstack{RFI \\($\alpha = 0.5$)} & \shortstack{RFI \\($\alpha = 1.0$)} \\ \hline \hline
\multirow{8}{*}{\rotatebox[origin=c]{270}{High}} & 
 \multirow{4}{*}{l} & 
 \multirow{2}{*}{l} & 
 l &
 305.451 &  \bf 5.027 &  9.165 &  - &  - &  - \\
\cline{4-10}
 & & & s &
 259.571 &  \bf 4.370 &  6.608 &  - &  - &  - \\
\cline{3-10}
 & & \multirow{2}{*}{s} & 
 l &
 8.821 &  \bf 0.740 &  1.974 &  15879.989 &  40814.085 &  - \\
\cline{4-10}
 & & & s &
 10.147 &  \bf 0.799 &  1.662 &  7212.395 &  17868.892 &  21866.164 \\
\cline{2-10}
 & \multirow{4}{*}{s} & 
 \multirow{2}{*}{l} & 
 l &
 3.050 &  \bf 0.280 &  1.741 &  - &  - &  - \\
\cline{4-10}
 & & & s &
 3.064 &  \bf 0.253 &  1.590 &  - &  - &  - \\
\cline{3-10}
 & & \multirow{2}{*}{s} & 
 l &
 0.290 &  \bf 0.096 &  0.505 &  869.717 &  1450.224 &  1720.670 \\
\cline{4-10}
 & & & s &
 0.287 &  \bf 0.077 &  0.578 &  434.343 &  713.357 &  650.564 \\
\cline{1-10}
\multirow{8}{*}{\rotatebox[origin=c]{270}{Low}} & 
 \multirow{4}{*}{l} & 
 \multirow{2}{*}{l} & 
 l &
 285.167 &  \bf 4.993 &  69.377 &  - &  - &  - \\
\cline{4-10}
 & & & s &
 256.525 &  \bf 4.432 &  458.153 &  - &  - &  - \\
\cline{3-10}
 & & \multirow{2}{*}{s} & 
 l &
 8.762 &  \bf 0.721 &  1.665 &  20763.900 &  24873.611 &  - \\
\cline{4-10}
 & & & s &
 10.156 &  \bf 0.720 &  4.135 &  8784.491 &  6108.177 &  27178.642 \\
\cline{2-10}
 & \multirow{4}{*}{s} & 
 \multirow{2}{*}{l} & 
 l &
 3.001 &  \bf 0.281 &  3.906 &  - &  - &  - \\
\cline{4-10}
 & & & s &
 3.061 &  \bf 0.284 &  40.593 &  - &  - &  - \\
\cline{3-10}
 & & \multirow{2}{*}{s} & 
 l &
 0.285 &  \bf 0.075 &  0.508 &  747.225 &  859.139 &  1610.464 \\
\cline{4-10}
 & & & s &
 0.307 &  \bf 0.085 &  0.752 &  361.877 &  586.522 &  522.050 \\
\cline{1-10}
\multirow{8}{*}{\rotatebox[origin=c]{270}{Zero}} & 
 \multirow{4}{*}{l} & 
 \multirow{2}{*}{l} & 
 l &
 287.191 &  \bf 4.898 &  - &  - &  - &  - \\
\cline{4-10}
 & & & s &
 259.578 &  \bf 4.350 &  - &  - &  - &  - \\
\cline{3-10}
 & & \multirow{2}{*}{s} & 
 l &
 8.714 &  \bf 0.737 &  6.995 &  24068.404 &  24868.802 &  45127.042 \\
\cline{4-10}
 & & & s &
 10.027 &  \bf 0.799 &  7.590 &  8136.108 &  6511.796 &  24328.727 \\
\cline{2-10}
 & \multirow{4}{*}{s} & 
 \multirow{2}{*}{l} & 
 l &
 3.006 &  \bf 0.289 &  965.906 &  - &  - &  - \\
\cline{4-10}
 & & & s &
 3.110 &  \bf 0.245 &  - &  - &  - &  - \\
\cline{3-10}
 & & \multirow{2}{*}{s} & 
 l &
 0.294 &  \bf 0.079 &  0.800 &  731.388 &  928.043 &  1204.162 \\
\cline{4-10}
 & & & s &
 0.294 &  \bf 0.091 &  1.260 &  309.829 &  547.799 &  669.768 \\
\cline{1-10}
\end{tabular}
\begin{tablenotes}
  \item '-': method either exceeds runtime limit (8 hours) or runs out of memory.
  \end{tablenotes}
\end{threeparttable}
\end{table}

\subsubsection{Runtime}
We measure the total wall-clock runtime of each data repairing method for all datasets. The results are shown in Table~\ref{tab:synruntime}. \system and GL are python based, non-parallelized programs, while RFI and PYRO are Java based, parallelized program. Since, most methods finish within hundreds of seconds, we limit the maximum runtime to eight hours. Overall, we see that \system's runtime is better than RFI's and \system has better column-wise scalability than both methods though poor row-wise scalability than PYRO.

\subsection{Performance on Real-World Data}
We evaluate the performance of all methods on the real-world datasets described in Section~\ref{exp_setup}. We first report the runtime of different methods and then present a qualitative analysis of the FDs they discover. A summary of our findings is shown in Table~\ref{tab:exp3_runtime}. We first focus on runtime. As shown both \system and PYRO can scale to large real-world noisy data instances. We see that \system only requires only 79 seconds to analyze a dataset with $\sim2$ million tuples and 18 attributes. As with the synthetic data RFI scales poorly. We next focus on the FDs discovered by the different methods.

\begin{table}[t]
\scriptsize
\center
\caption{Quantitive Results over Real-world Datasets}
\label{tab:exp3_runtime}
\begin{tabular}{c |l c c | c c c c}
Dataset & & \system & GL & PYRO & RFI(.3) & RFI(.5) & RFI(1.0)\\ \hline \hline
\multirow{2}{*}{Hospital} 
& runtime (sec) & \bf 0.318 & - & 1.029 & 3249.8 & 10272.8 & 17712.8\\
& \# of FDs & 9 & - & 434 & 16 & 16 & 16\\
\hline
\multirow{2}{*}{Food} 
 & runtime (sec) & 14.433 & \bf 0.924 & 5.059 & - & - & - \\
  & \# of FDs & 11 & 16 & 156 & - & - & - \\
  \hline
\multirow{2}{*}{Physician} 
 & runtime (sec) & 79.068 & \bf 5.920 & 55.978 & - & - & -\\
   & \# of FDs & 4 & 6 & 528 & - & - & - \\
\hline
\end{tabular}
\begin{tablenotes}
  \item '-' for GL: too few data samples makes the matrix to ill-conditioned to solve
  \item '-' for RFI: did not complete within eight hours.
  \item * this experiment was executed on a different machine with 4 CPUs (each is a 20-core Intel(R) Xeon(R) Gold 6148 with hyper-threading), 0.5TB RAM
 \end{tablenotes}
\end{table}

\begin{figure}
\includegraphics[width=0.35\textwidth]{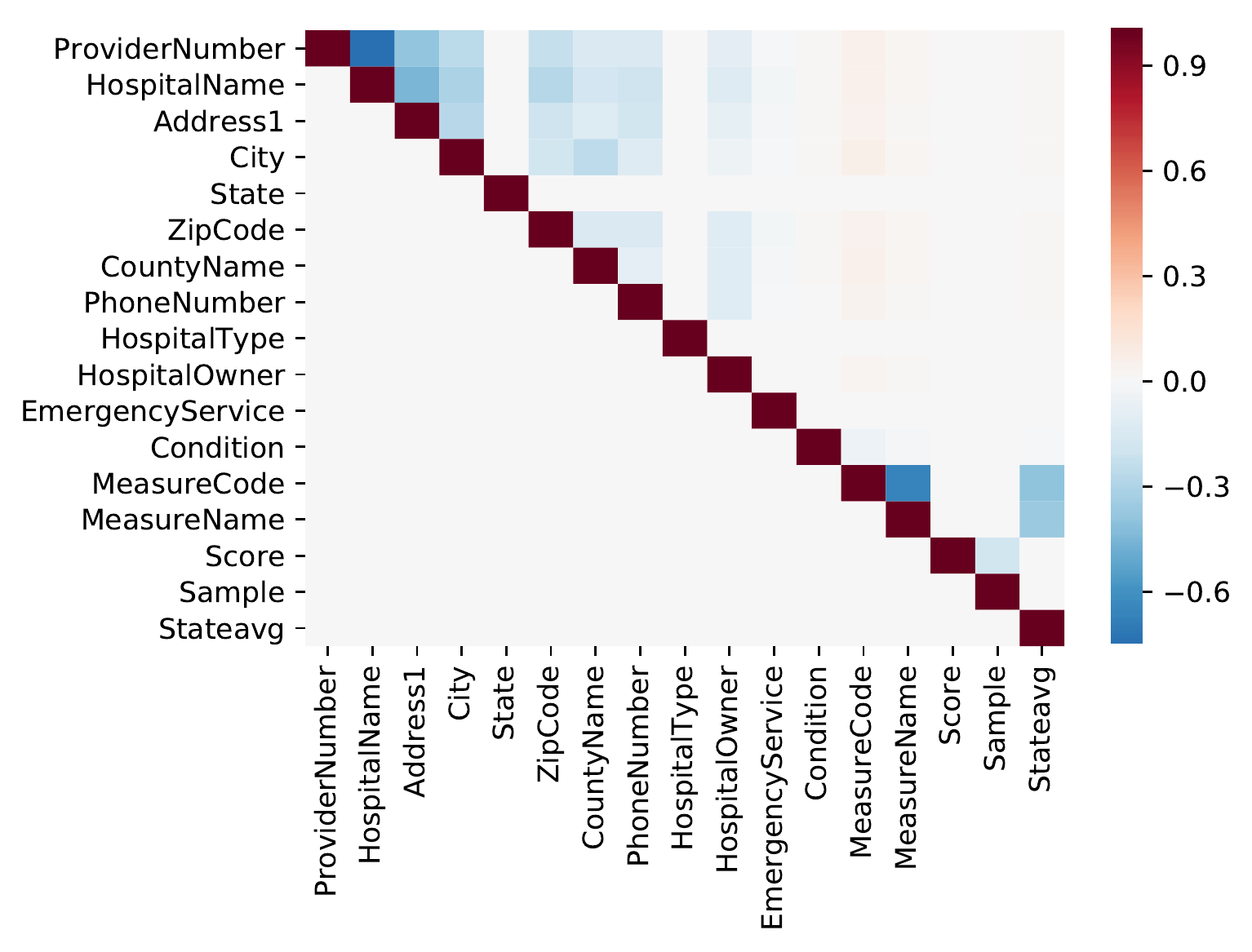}
\caption{The autoregression matrix estimated by \system for the Hospital dataset.}
\label{fig:gld_hospital}
\end{figure}

We see that \system, GL, and RFI find a number of FDs that is always less than the number of attributes in the input dataset. On the other hand, PYRO finds hundreds of FDs for each dataset. These results are consistent with the FD interpretation adopted by each system~\ref{sec:prelims}. We now analyze some of the FDs discovered different systems. We focus on the FDs discovered for Hospital. We consider the FDs discovered by \system. A heatmap of the regression matrix of \system's model is shown in Figure~\ref{fig:gld_hospital}. We find that the discovered FDs are meaningful. For example, we see that attributes `Provider Number' and `Hospital Name' determine most other attributes. We also see that `Address1' determines location-related attributes such as `City', `Zip code' and `County'. We also find that attribute `Measure Code' determines `Measure Name' and that they both determine `StateAvg'. In fact, `StateAvg' corresponds to the concatenation of the `State', and `Measure Code' attributes. The reader may wonder why the `State' attribute is found to be independent of every other attribute. We attribute this to the fact that hospital dataset only contains two states with one appearing nearly 89\% of time. Enforcing a sparse structure, \system weakens the role of `State' in deterministic relations. These results show that \system can identify meaningful FDs in real-world datasets.

\begin{figure}
\includegraphics[width=0.4\textwidth]{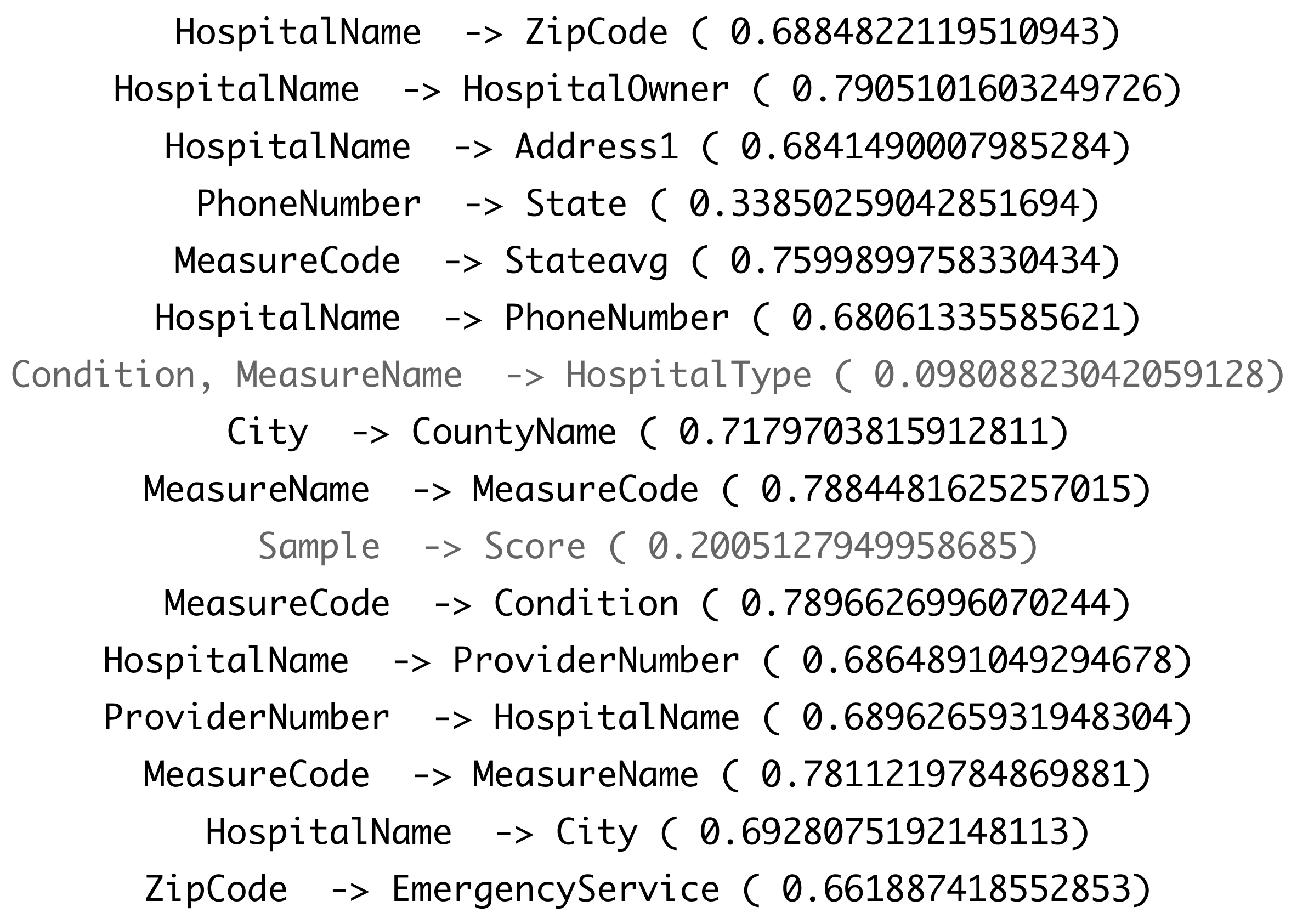}
\caption{The FDs discovered by RFI for Hospital.}
\label{fig:ri_hospital}
\end{figure}

We consider the competing methods. For RFI, the results are consistent across all three alphas, so we pick the one with highest alpha (lower approximate rate). RFI outputs 18 FDs that are shown in Figure~\ref{fig:ri_hospital}. The value in the parenthesis is the reliable fraction of information, the score proposed by RFI to select AFDs. After eliminating FDs with low score, we find that most of FDs discovered by RFI are also meaningful. However, it has the problem of overfitting to the dataset. Specifically, for the FD `ZipCode' $\rightarrow$ `EmergencyService', this relation holds for the given dataset instance, but does not convey any real-world meaning. We attribute this behavior to the fact that the domain of `ZipCode' is really large while `Emergency Service' only has a binary domain. This makes it more likely to observe a spurious FD when the number of data samples is limited. This finding matches RFI's performance for the synthetic datasets. For PYRO, we find that it discovers hundreds of FDs that are not particularly meaningful for data preparation tasks. For instance, PYRO finds 24 FDs that determine the attribute `Address1'.

\subsection{Using \system to Automate Data Cleaning} Recent work~\cite{hc} showed that integrity constraints such as FDs can be used to train machine learning models for data cleaning in a weakly supervised manner. A limitation of this work is that it relies on users to specify these constraints. Here, we test if \system can be used to automate this process and address this pain point. For our experiments, we use the open-source version of the system from~\cite{hc}, as it provides a collection of manually specified FDs for the Hospital dataset. We perform the following experiment: we compare the manual FDs in that repository with the FDs discovered by \system. The precision, recall, and $F_1$ reported by the data cleaning system for the manual constraints are 0.91, 0.70, and 0.79 respectively, while the corresponding metrics for the FDs discovered by \system is 0.93, 0.72, and 0.81. We se that this performance is comparable to the manually specified FDs, thus, providing evidence on the applicability of \system to discover FDs that are useful in downstream data preparation tasks.

\subsection{Micro-benchmark Results}
Finally, we report micro-benchmarking results: (1) we evaluate the scalability of \system and demonstrate its quadratic computational complexity with respect to number of attributes; (2) evaluate the effect of increasing noise rates on the performance of \system.

\subsubsection{Column-wise Scalability}
\begin{figure}
\includegraphics[width=0.38\textwidth]{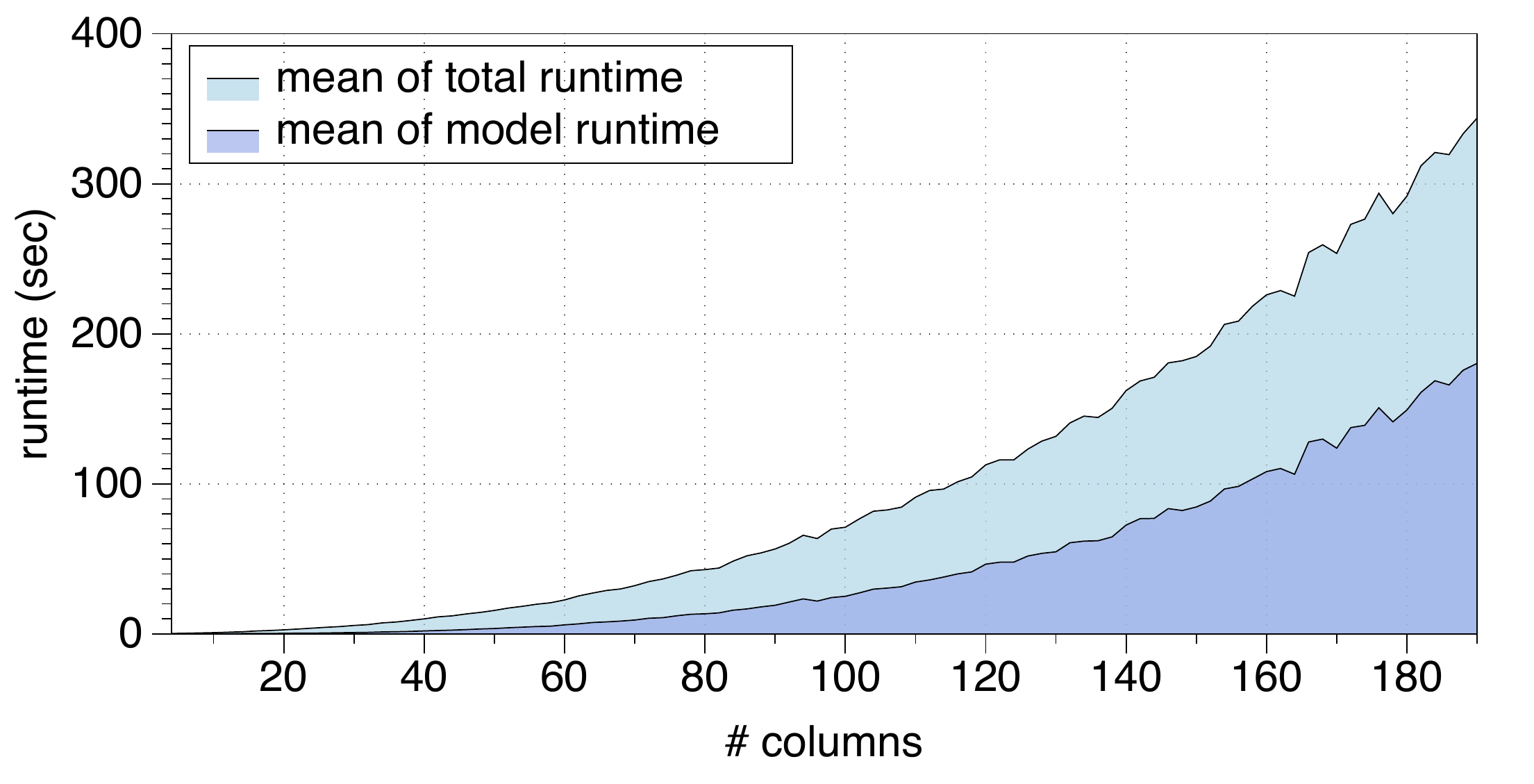}
\caption{Columns-wise Scalability of \system.}
\label{fig:exp4-2}
\end{figure}

Based on our discussion in Section~\ref{sec:model}, \system exhibits quadratic complexity instead of exponential complexity with respect to the number of columns in a dataset. We experimentally demonstrate \system's scalability. We generate a collection of synthetic datasets where we keep all settings fixed except for the number of attributes, which we range from 4 to 190 with a increase step of two. For each number of columns, we generate five datasets and calculate the average runtime for each columns size. In addition, we log both the total runtime (including data loading and data transformation) and the structure learning runtime. The results are shown in Figure~\ref{fig:exp4-2} and validate the quadratic scalability of \system as the number of attributes increase. 

\subsubsection{Effect of Increasing Noise Rates}
\begin{figure}
\includegraphics[width=0.4\textwidth]{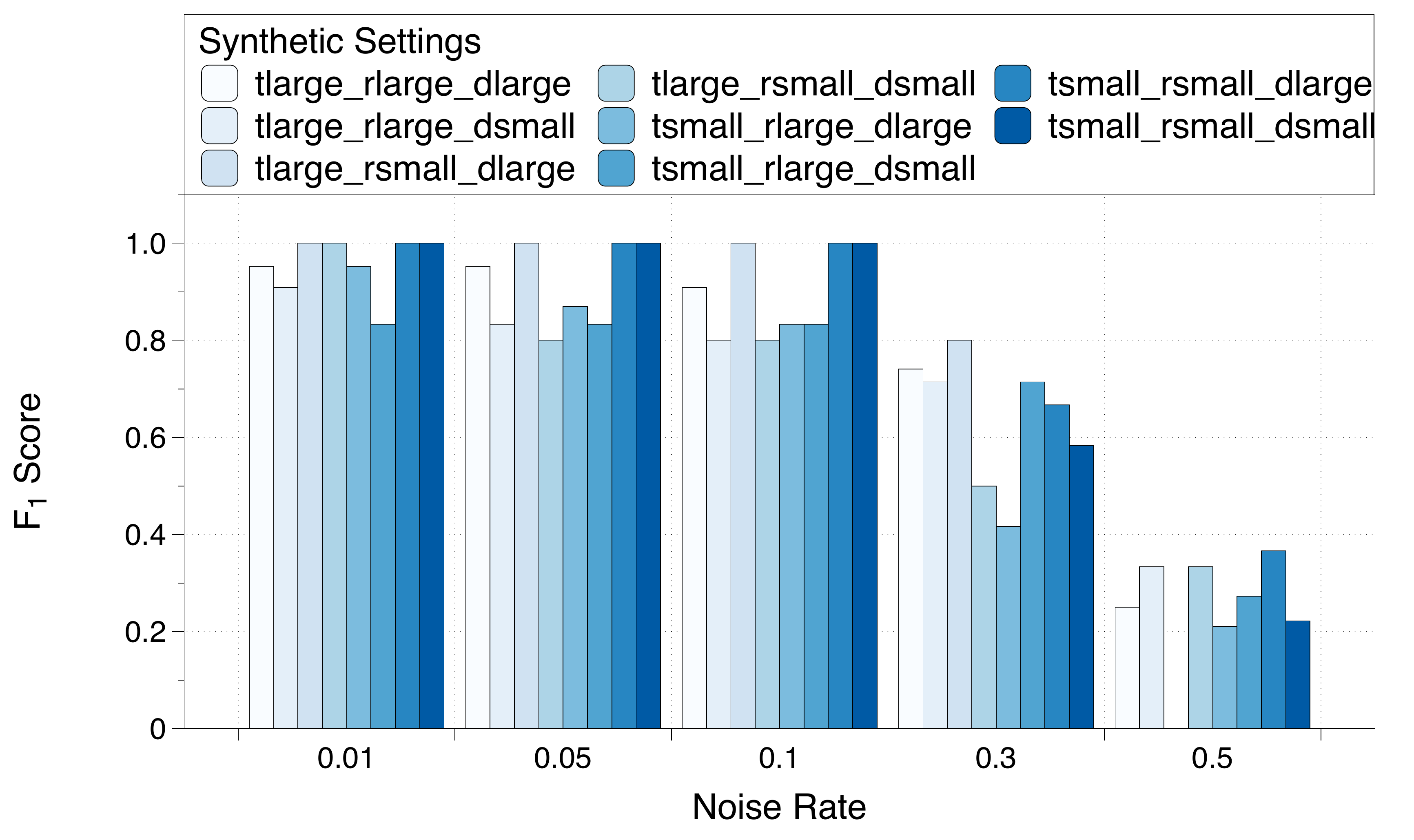}
\caption{Effect of Increasing Noise Rates. Dataset names correspond to the setting that was used (see Table~\ref{tab:synthetic_settings}).}
\label{fig:exp4-4}
\vspace{-10pt}
\end{figure}

In this experiment, we evaluate how \system performs as the noise rate increases. For this experiment we generate a new set of synthetic datasets that is different from that of Section~\ref{sec:end2end}. Again we generate five instances per dataset setting (see Table~\ref{tab:synthetic_settings} for our settings) and measure the performance of \system for noise rates in $\{0.01, 0.1, 0.3, 0.5\}$. We report the median $F_1$ score in Figure~\ref{fig:exp4-4}. As expected, the performance of \system deteriorates as the noise increases, however, \system is shown to be robust to high error rates.

\section{Conclusions}\label{sec:conclusions}
We introduced \system, a structure learning framework to solve the problem of FD discovery in relational data. A key result in our work is to model the distribution that FDs impose over pairs of records instead of the joint distribution over the attribute-values of the input dataset. Specifically, we introduce a method that convert FD discovery to a structure learning problem over a linear structured equation model. We empirically show that \system outperforms state-of-the-art FD discovery methods and can produce meaningful FDs that are useful for downstream data preparation tasks.

\bibliographystyle{ACM-Reference-Format}
\bibliography{profiler-bibliography.bib}

\end{document}